\documentclass[10pt]{article}
\usepackage{amssymb,amsmath,mathtools}
\usepackage{authblk}

\newcommand{\rd}{\mathrm{d}}

\newcommand{\td}[2]{\frac{\rd #1}{\rd #2}}

\newcommand{\eps}{\epsilon}

\newcommand{\beq}{\begin{equation}}
\newcommand{\eeq}{\end{equation}}

\newcommand{\ii}{\mathrm{i}}

\newcommand{\ph}{\varphi}

\advance\hoffset by -1.1truecm
\advance\voffset by -1.0truecm

\begin{document}
\title{Dispersion relations and bending losses of cylindrical and spherical shells, slabs, and slot waveguides}

\author{Gregory Kozyreff$^{\footnote{gkozyref@ulb.ac.be}}$ and Nirmalendu Acharyya}

\affil{\small{Optique Nonlin\'eaire Th\'eorique, Universit\'e libre de Bruxelles (U.L.B.), CP 231, Campus de la Plaine, 1050 Bruxelles, Belgium}}

\date{}
\maketitle 


\begin{abstract}
We derive formulas for Whispering Gallery Mode resonances and bending losses in infinite cylindrical dielectric shells and sets of concentric cylindrical shells. The formulas also apply to spherical shells and to sections of bent waveguides. The derivation is based on a WKB treatment of Helmholtz equation and can in principle be extended to any number of concentric shells. A distinctive limit analytically arises in the analysis when two shells are brought at very close distance to one another. In that limit, the two shells act as a slot waveguide. If the two shells are sufficiently apart, we identify a structural resonance between the individual shells, which can either lead to a substantial enhancement or suppression of radiation losses.
\end{abstract}




\section{Introduction}

The quality factor, $Q$, of whispering gallery mode (WGM) resonators is ultimately limited by  material purity and radiation losses. The former seems to be nearly attained in large crystalline resonators, where $Q>10^{11}$ has been demonstrated~\cite{Savchenkov-2007b}. The latter, equally unavoidable, is known to decrease exponentially with the cavity radius, yet a simple, explicit formula is lacking. A precise analytical knowledge of the radiation losses could be useful, as it could indicate ways of keeping a large $Q$ while reducing the cavity size. 

In the presence of spherical or cylindrical symmetry, the characteristic equation giving the complex WGM frequencies generally involves Bessel function of both large orders and large arguments. This double limit is a delicate one to handle numerically, which motivates one to use asymptotic representations. For full and uniform dielectric spheres or cylinders, the positions of the resonances are most accurately given in Schiller's paper~\cite{Schiller-1993}, who extended the analysis of Lam, Leung and Young~\cite{Lam-1992}.  These authors also successfully computed the radiative linewidth by substituting their result in the argument of an appropriate spherical Bessel function. However, this procedure requires one to employ Debye's expansion~\cite{Abramowitz} of Bessel functions, which is somewhat fastidious and not particularly illuminating. As a result, therefore, one usually keeps solving the WGM characteristic equation numerically to assess the radiation losses \cite{Prkna-2004,Spillane-2005,Agha-2006,Dumeige-2006}.

Meanwhile, there is a strong and related interest to assess bending losses in classical optical waveguides and slot waveguides. Indeed, miniaturization of photonic integrated circuits (PIC) drives the design towards sharper bends and towards smaller ring or stadium cavities. It would then be useful to simply estimate the rise in radiative losses associated to this trend. A formula already exists for bent monomode fibers, which is due to Chang and Barnes~\cite{Chang-1973} and Marcuse~\cite{Marcuse-1976} and which has recently been improved by analytical and numerical investigations~\cite{Hiremath-2005,Jedidi-2005,Schermer-2007}. However, Marcuse's formula is only valid in the limit where the refractive index in the guiding is very close to that of the surroundings. If the difference in refractive indexes is not small, as is usually the case with PIC, the present analysis takes over.

Our study also finds application in WGM sensing, where the high $Q$ of micro-rings or micro-spheres promises highly sensitive label-free detection of molecules.  Recently, it has been shown numerically that to surround a micro-ring with an external bus waveguide could increase the quality factor \cite{Cai-2015}. In the same vein, concentric micro-rings have been proposed as an optical improvement over simple micro-rings \cite{Li-2009,Malmir-2016}. We address this question in this paper by considering concentric shells, which can be viewed as analytically tractable models for rings. We show analytically how the $Q$-factor can indeed be increased but, also, drastically lowered by an external ring, depending on the geometrical parameters.

Various methods can be used to study WGM. Their frequencies can in principle be directly found as the roots of the characteristic equation. Such characteristic equation may be written with no difficulty if the geometry is a set of concentric cylindrical layers, by solving Helmholtz equation in terms of Bessel functions and applying proper boundary conditions. Previous analyses \cite{Lam-1992,Schiller-1993} are based on an asymptotic study of the characteristic equation. In the present paper, we apply the WKB method on the Helmholtz equation itself. This approach yields not only the complex frequencies of WGM but also their spatial distribution. Our analysis is close in spirit to that performed by Chang and Barnes~\cite{Chang-1973}, except that we do not make the assumption of a very weak index contrast and that we don't stretch the radial coordinate to obtain a Schr\"odinger equation. More recently, the method of Keller and Rubinow~\cite{Keller-1960} was used to compute the position of WGM resonances in spheroidal cavities \cite{Gorodetsky-2006,Gorodetsky-2007,Demchenko-2013}. However, this method was devised for bounded domain and it is therefore not clear if it can be applied to compute radiative losses.

Before stating our results, let us recall that the treatments of a cylindrical cavity of radius $R$ and of a segment of bent waveguide are locally equivalent. In the former, the field has the azimuthal and temporal dependence
\begin{align}
\exp \ii\left(\ell\ph-kc t\right), &&k=k_r-\ii k_i,
\label{azimuthal}
\end{align}
whereas in the latter, this spatio-temporal dependance is replaced by
\begin{align}
\exp \left[\ii\left(\beta s-k_r c t\right) - \alpha s\right]
&&\beta&=\ell/R,
&\alpha&=k_i\ell/(k_r R),
\end{align}
where $s$ is the arc length along the waveguide. Hence, the dispersion relation $\beta=\beta(k_r)$ and attenuation coefficient $\alpha$ may directly be deduced from the relations $k_r=k_r(\ell)$, $k_i=k_i(\ell)$.

We also remind that the solutions of Helmholtz equation on spheres and on cylinders can be mapped onto one other and are therefore equivalent~\cite{Jackson-1999}. The synthesis in the dispersion formulas is achieved through the parameter
\beq
\nu=\left\{
\begin{matrix*}[l]
\ell& \text{for cylindrical geometries,}\\
\ell+1/2& \text{for spherical geometries.}
\end{matrix*}
\right.
\eeq
Mathematically, the defining feature of WGM is that $\nu\gg1$~\cite{Kozyreff-2011}.

All formulas to follow are for materials of refractive index $n$ embedded in air. For materials of refractive index $n_2$ in an environment of refractive index $n_1$, $n$ should be understood as  the ratio $n_2/n_1$, while $c$ should be changed into $c/n_1$. 

\subsection{Summary of the results}

In terms of the normalised real frequency
\beq
x =\frac{k_rR}\nu,
\eeq
we find that the radiative decay rate for full spheres and cylinders is given by
\begin{align}
k_iR=\Gamma_0&\sim
\frac{2  e^{2\nu S(x)} }{\xi  \sqrt{n^2-1}},
&S(x)&=\sqrt{1-x^2}-\ln\frac{1+\sqrt{1-x^2}}{x},
\label{loss_shell}
\end{align}
where $\xi=1$ for TE modes and $\xi=n^2$ for TM modes. This formula is, to our knowledge, new, although it could be inferred from consulting Lam {\it et al.}'s paper~\cite{Lam-1992} alone. We have checked its numerical validity.

For more general circular geometries, the radiative losses are
\beq
k_i R = \sigma\times\Gamma_0,
\label{shapefactor}
\eeq
and our main contribution is to provide the {\it shape factor} $\sigma$ for shells in Eq.~(\ref{shape:shell}), for concentric shells in Eqs.~(\ref{shape:exterior}) and (\ref{shape:interior}), and for slots in Eq.~(\ref{shape:slot}).

Our analysis also yields the resonances $x$. Closed WGM trajectories must contain an integer number, $\ell$, of wavelengths, so $x\approx1/n$. More generally,
\begin{align}
x&\sim\frac{1+\eps \tau}n+O\left(\eps^{3/2}\right),
&\eps&=2^{-1/3}\nu^{-2/3}.
\label{epsilon:1}
\end{align}
In the case of full spheres and cylinders~\cite{Lam-1992}, it is well-known that $\tau$ is a root of $Ai(-z)$, but for shells of thickness $h$, $\tau$ must solve
\begin{align}
\mathcal{D}(\tau,\hat h)&=0,
&\hat h&=\eps^{-1}(h/R),
\end{align}
where the characteristic function is
\beq
\mathcal{D}(\tau,\hat h)\equiv Ai(-\tau)Bi(\hat h-\tau)-Ai(\hat h-\tau) Bi(-\tau).
\label{eq:tau}
\eeq
The roots $\tau$ of the above characteristic equation yield the first correction to the leading-order expression of the annular resonance. An extended asymptotic expansion of $x$ with $O(\nu^{-5/3})$ accuracy is provided in Eq.~(\ref{res_shell}). This expression for $x$ also holds for sets of concentric shells, if they are sufficiently apart. 

If two shells are brought at very short distance from one another, then guiding can take place within the narrow low-index region for the TM field. The low-index  gap thus forms a slot waveguide, as in~\cite{Almeida-2004,Chao-2007,Kargar-2011}. How narrow must the gap be for this to happen? Defining the low-index region as $R_1<r<R_2$ we find that this dynamical regime, in the large-$\nu$ limit, corresponds to $k^{4/3}R_1^{1/3}(R_2-R_1)=O(1)$. At telecom wavelengths, this estimates yields a few tens of nm, consistently with previous studies. The slot resonances are given by
\begin{align}
\mathcal{D}(\tau,\hat h_1+\hat h_2)+\pi\Delta \mathcal{D}(\tau, \hat h_2)\mathcal{D}(\tau-\hat h_1,\hat h_1 )&=0,
&\Delta= 2^{-1/3}\nu^{4/3}\left(n^2-1\right)\frac{R_2-R_1}{R_1},
\end{align}
where $\hat h_{1,2}=h_i/(\eps R_i)$ are the reduced thicknesses of the high-index shells that enclose the slot.

The rest of the paper is organised as follows. In Sec.~\ref{sec:shell}, we develop the WKB analysis and treat the single shell or bent slab waveguide. In Sec.~\ref{sec:coupled}, we analyse two weakly coupled concentric shells or waveguides. In Sec.~\ref{sec:slot}, we treat the slot-waveguides. Finally, we discuss our results and conclude.

\section{Simple shells and bent slabs}\label{sec:shell}
Given the azimuthal dependance in (\ref{azimuthal}), Maxwell's equation lead to
\begin{align}
\td{^2\psi}{r^2}+\frac1r\td\psi{r}+\left(n_i^2k^2-\frac{\nu^2}{r^2}\right)\psi&=0,
&n_i&=\left\{
\begin{matrix*}[l]
n& R-h<r<R,\\
1& \text{elsewhere}
\end{matrix*}
\right.
\label{Helmholtz}
\end{align}
with the usual radiation condition $\psi\sim\exp(\ii kr)/\sqrt{r}$ as $r\to\infty$ and the continuity condition
\beq
\left[\psi\right]_-^+=\left[\frac{1}{\xi}\td\psi {r}\right]_-^+=0
\label{eq:cont1}
\eeq
at interfaces, where $\xi=1, \psi= E_z$ for TE waves and $\xi=n_i^2,\psi= H_z$ for TM waves. The first step in our analysis is to rescale the radial coordinate as
\beq
y=k r/\nu,
\label{def:y}
\eeq
which transforms Eq.~(\ref{Helmholtz}) into
\begin{align}
\td{^2\psi}{y^2}+\frac1{y}\td\psi{y}&=\nu^2q(y)\psi,
&q(y)&=
\left\{
\begin{matrix*}[l]
y^{-2}-n^2&  x(1-h/R)<y<x,\\
y^{-2}-1& \text{elsewhere,}
\end{matrix*}
\right.
\label{Helm:1}
\end{align}
subjected to conditions~(\ref{eq:cont1}), with ``$\rd/\rd r$" replaced by ``$\rd/\rd y$''. This sets the problem  in a suitable form to apply the WKB technique in the large-$\nu$ limit~\cite{Benderbook}. Note that the eigenvalue $k$ or, equivalently, $x$, is moved from the differential equation to the location of the boundary condition in the $y$ variable. In what follows, we will successively solve Eq.~(\ref{Helm:1}) outside and inside the shell. As usual, the characteristic equation will then emerge from the continuity conditions.

\subsection{Outside the shell}\label{sec:outside}
Wherever $q(y)$ is different from zero, a WKB analysis in Appendix A yields 
\beq
\psi_\pm\sim\text{const.}\times\frac{e^{\pm\nu S(y) } }{\left|1-y^2\right|^{1/4}}\left(1+\sum_{j\ge1} \nu^{-j}f_j\right),
\label{WKB}
\eeq
with
\begin{align}
f_1&=-\frac{2+3y^2}{24\left(1-y^2\right)^{3/2}},&
f_2&=\frac{4+300y^2+81y^4}{1152\left(1-y^2\right)^{3}},\quad\dots
\end{align}
and where
\beq
S(y)=
\left\{
\begin{matrix*}[c]
\ii\left(\sqrt{y^2-1}-\arccos(1/y)\right)&  y>1,\\
\sqrt{1-y^2}- \ln\frac{1+\sqrt{1-y^2}}{y}& y<1.
\end{matrix*}
\right.
\eeq
From the form of $S(y)$ above, one sees that $y>1$ is the \emph{radiation zone}, where the field progressively acquires the form of a radially outgoing wave as $y$ becomes large. Conversely, for smaller radii, $y<1$, the field has an exponential dependance on the radius. We therefore call this region the \emph{evanescent zone}. The near field outside the shell is
\begin{align}
\psi&\sim\frac{Ae^{ \nu S(y) }+Be^{-\nu S(y) }}{\left|1-y^2\right|^{1/4}}\left(1+\sum_{j\ge1} \nu^{-j}f_j\right),
&x<y<1
\label{exterior}
\end{align}
while the radiation condition in the far field imposes
\begin{align}
\psi&\sim
\frac{2Ce^{\ii\pi/4} e^{ \nu S(y)}
}
{\left(y^2-1\right)^{1/4}}
\left(1+\sum_{j\ge1}\nu^{-j}f_j\right),
& y>1.
\label{radzone}
\end{align}
The connection between the constants $A$, $B$ and $C$ above can be found by a local analysis in the vicinity of $y=1$, as explained  in~\cite{Benderbook}. Briefly, in that region, we find that the solution is locally given by
\begin{align}
\psi&\sim 2^{5/6}\pi^{1/2}\nu^{1/6}C\left[\ii \, Ai\left(-w\right)+  Bi\left(-w\right)\right],
&w&=\left(y-1\right)/\eps,
\label{turning}
\end{align}
where $Ai(z),Bi(z)$ are Airy's functions~\cite{Abramowitz} and $\eps$ was defined in (\ref{epsilon:1}). That particular combination ensures that the correct far field~(\ref{radzone}) be recovered as $w\to\infty$. Conversely, matching (\ref{turning}) with~(\ref{exterior}) as $w\to-\infty$ yields
\begin{align}
A&=\ii C,&B&=C.
\end{align}
Hence, in the evanescent zone at the exterior of the shell, we have
\begin{align}
\psi&\sim C\frac{\ii e^{\nu S(y) }+e^{-\nu S(y) }}{\left|1-y^2\right|^{1/4}}\left(1+\sum_{j\ge1} \nu^{-j}f_j\right),
&x<y<1.
\label{exterior2}
\end{align}
In this last expression, the  term $\ii \exp(\nu S)$ is exponentially small in the vicinity of the shell but is also exponentially growing away from it. In their asymptotic description of WGM of dielectric spheres, Little {\it et al.} omit this contribution, together with the radiating field very far away from the sphere~\cite{Little-1999}. While the error incurred on the position of the resonances is negligible, it is this term that allows one to properly determine the radiation losses.

On the other side of the shell, we have
\begin{align}
\psi&\sim\frac{Ge^{\nu S(y) } }{\left|1-y^2\right|^{1/4}}\left(1+\sum_{j\ge1} \nu^{-j}f_j\right),
&y<x\left(1-h/R\right)
\label{interior}
\end{align}
to ensure that the field stay bounded (note that $S(y)<0$ for $y<1$.) 

\subsection{Inside the shell}
\begin{figure}
\centering
\includegraphics[width=7.5cm]{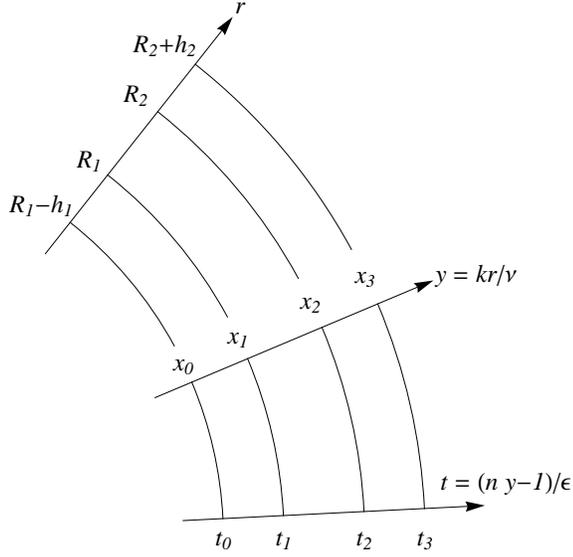}
\caption{The unscaled and normalised radial coordinates used in our analysis. For the one-shell problem, $R_1=R$, $h_1=h$.}
\label{fig:coord}
\end{figure}

Given that $h/R\ll1$, we have $y\approx x$ throughout the guiding region. Moreover, it is well known that the leading-order approximation for WGM resonances is $nkR\approx \nu$, {\it i.e.} that $x\approx1/n$. Hence, in the guiding region we rescale $y$ as 
\begin{align}
y&= \frac{1+\eps t}n,
&\text{(see Fig.~\ref{fig:coord})}
\label{def:t}
\end{align}
 where $\eps$  is the small parameter defined in (\ref{epsilon:1}). 
The equation for $\psi$ then becomes
\begin{align}
\psi&=F\left(t\right),
&F''(t)+tF(t)\sim\eps\left(\frac32t^2F(t)-F'(t)\right) +O(\eps^2).
\end{align}
In the limit $\eps\to0$, one recognises above Airy's equation, with independent solutions $Ai(-t),Bi(-t)$. To obtain a solution that is correct to $O(\eps)$ inclusive, one may treat the above equation as a standard perturbation problem or, in a more expeditious way, we may seek a solution of the form $\left(1+ \eps a t+\ldots\right)Ai\left(-t+\eps(b t+ ct^2+\ldots)\right)$ and determine the constants $a,b,c,\ldots$ so as to solve the equation up to the desired order.
We find the two solutions
\begin{align}
\mathcal{A}(t)&\sim\left(1-\frac{\eps t}5\right)Ai\left(-t+\frac{3\eps t^2}{10}\right),
&\mathcal{B}(t)&\sim\left(1-\frac{\eps t}5\right)Bi\left(-t+\frac{3\eps t^2}{10}\right).
\end{align}
In terms of these:
\begin{align}
\psi &= H \mathcal{A}(t) + I \mathcal{B}(t),\\
\td\psi{y}&=\frac{n}\eps \left[H \mathcal{A}'(t) + I \mathcal{B}'(t)\right]
\sim -n\nu 
\left[\frac{H Ai'\left(-t\right)+ I Bi'(-t)}{2^{-1/3}\nu^{1/3}}+O\left(\nu^{-1}\right) \right].
\end{align}

\subsection{Continuity and characteristic equation}
\begin{figure}
\centering
\includegraphics[width=8cm]{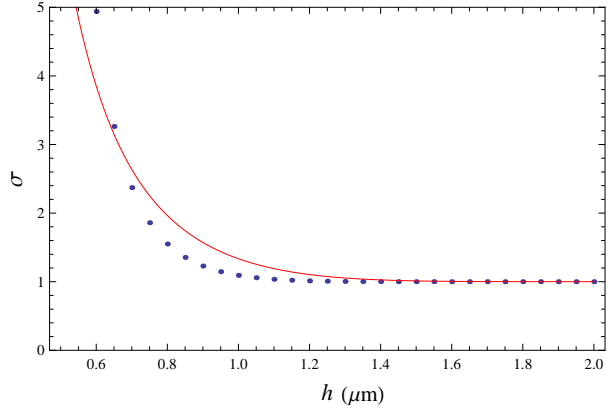}
\caption{Shape factor for a single annular cylinder, with outer radius $R=6.4\mu$m and varying thickness $h$. Refractive index $n=1.65$, orbital number $\nu=60$. Solid line: analytical formula. Dots: numerical calculation.}
\label{fig:1ring}
\end{figure}

As discussed earlier, we aim to find $x$, or, equivalently $t=t_1$, where
\beq
x=\frac{kR}\nu= \frac{1+\eps t_1}n.
\eeq
The normalised frequency $x$ is also the location of the outer radius of the shell in the $y$ coordinate, while $t_1$ is the location of the same boundary in the $t$ coordinate. The inner radius is at $y=x_0= x(1-h/R)$ (see Fig.~\ref{fig:coord}). Let us introduce the reduced thickness
\beq
\hat h = \frac{h}{\eps R},
\eeq
which we assume to be $O(1)$. Then the value of $t$ at the inner radius is $t_0= t_1-(1+\eps t_1) \hat h$. The continuity relations~(\ref{eq:cont1}) are, up to $O(\nu^{-1})$ corrections,
\begin{align}
H \mathcal{A}(t_1) + I \mathcal{B}(t_1)&=C\frac{\ii e^{\nu S(x) }+e^{-\nu S(x) }}{\left|1-x^2\right|^{1/4}} ,
\label{contrel1}\\
H \mathcal{A}(t_0) + I \mathcal{B}(t_0)&=\frac{Ge^{\nu S(x_0) } }{\left|1-x_0^2\right|^{1/4}},\\
\frac{-n  }{\xi}
\left[\frac{H Ai'\left(-t_1\right)+ I Bi'(-t_1)}{2^{-1/3}\nu^{1/3}}\right]
&=C \sqrt{q(x)}\frac{\ii e^{\nu S(x) }-e^{-\nu S(x) }}{\left|1-x^2\right|^{1/4}},
\label{C}\\
\frac{-n }{\xi} \frac{H Ai'\left(-t_0\right)+ I Bi'(-t_0)}{2^{-1/3}\nu^{1/3}}
&=G \sqrt{q(x_0)}\frac{e^{\nu S(x_0) } }{\left|1-x_0^2\right|^{1/4}}.
\label{G}
\end{align}
Note that, from the assumption that $\hat h=O(1)$, $x_0$ differs from $x$ by an $O(\nu^{-2/3})$ quantity only. Hence, 
\beq
q(x),q(x_0)\sim n^{2}-1+O\left(\nu^{-2/3}\right)
\eeq
Using Eqs.~(\ref{C}) and (\ref{G}) to eliminate  $C$ and $G$, we eventually obtain:
\begin{align}
H \mathcal{A}(t_1) + I \mathcal{B}(t_1)&\sim\alpha\sqrt{\eps} \left[ H Ai'\left(-t_1\right)+ I Bi'(-t_1)\right]\left(1+2\ii e^{2\nu S(x)}\right) ,\label{shell1}\\
H \mathcal{A}(t_0) + I \mathcal{B}(t_0)&\sim-\alpha\sqrt{\eps}  \left[H Ai'\left(-t_0\right)+ I Bi'(-t_0) \right],
\label{shell2}
\end{align}
correct up to $O(\nu^{-1},e^{4\nu S(x)})$ terms, where we define, for brevity,
\beq
\alpha=\frac{\sqrt{2}n}{\xi\sqrt{n^2-1}}.
\label{def:alpha}
\eeq
It is immediate  from Eqs.~(\ref{shell1}) and (\ref{shell2}) that, in the limit  $\eps\to0$,  $t_1\to\tau$, where
\beq
\mathcal{D}(\tau,\hat h) \equiv 
Ai(-\tau )Bi(\hat h -\tau )- Ai(\hat h -\tau ) Bi(-\tau )=0.
\eeq
A more systematic asymptotic resolution, outlined in Appendix B, eventually yields
\begin{multline}
x\sim  \frac{1+2^{-1/3}\nu^{-2/3}\tau}n
-\frac{\nu^{-1} }{\xi \sqrt{ n^2-1}} \frac{  1+P^2} {1 -P^2} 
+\frac{2^{-2/3}\nu^{-4/3}}{n}\times\\
\Bigg[
\frac{3\tau^2}{10}
-\frac{P^2}{1-P^2}\left(\frac{3\hat h^2+4\hat h\tau}{10} 
-
\frac{8n^2}{\xi^2(n^2-1)}
\frac{\left(1-P\tilde P\right)Bi'(\hat h-\tau)}{\left(1-P^2\right)^2 Bi(\hat h-\tau)}
\right)
\Bigg]
+O\left(\nu^{-5/3}\right),
\label{res_shell}
\end{multline}
where
\begin{align}
P&=\frac{Bi(-\tau)}{Bi(\hat h -\tau)} , &\tilde P&=\frac{Bi'(-\tau)}{Bi'(\hat h -\tau)}.
\end{align}
In addition, an exponentially small imaginary contribution is found in $x$, which leads to Eq.~(\ref{shapefactor}), with the shape factor
\beq
\sigma = \frac1{1-P^2},
\label{shape:shell}
\eeq
Note that the numbers $P$ and $\tilde P$ appearing in the above expressions rapidly become small as $\hat h -\tau$ increases above zero. In that limit, the formula's in Ref.~\cite{Lam-1992} are recovered. This quantitatively indicates how the shell becomes optically equivalent to a full cylinder with increasing thickness. To check the validity of Eq.~(\ref{shape:shell}), we solved the exact characteristic equation of a full cylinder and of a shell having the same outer radius but varying thickness $h$ (see Appendix C). The ratio of the imaginary parts of the complex roots provides a numerical value of $\sigma$. Due to numerical limitations, we took $\nu= 60$, meaning a vacuum wavelength of $\pm1\mu$m for an outer radius of $6.4\mu$m. The comparison in Fig.~\ref{fig:1ring} shows good agreement with the analytical formula, given the moderately large value of $\nu$.

\section{Coupled shells}\label{sec:coupled}
Let us now consider two concentric annuli,  delimited by $R_1-h_1<r<R_1$ and $R_2<r<R_2+h_2$, with $R_2>R_1$ and with refractive index $n$. In the $t$-variable, the four interfaces are located at $t_0<t_1<t_2<t_3$ and we define $x_{j}=kR_i/\nu$, $j=1,2$, see Fig.~\ref{fig:coord}. From the definitions~(\ref{def:y}) and~(\ref{def:t}) of the $y$  and $t$ variables, we have
\begin{align}
t_3-t_2&\sim\frac{h_2}{\eps R_2}\equiv\hat h_2
&\text{and}&&t_1-t_0&\sim\frac{h_1}{\eps R_1}\equiv\hat h_1.
\end{align}
The interior and exterior fields are the same as in Sec.~\ref{sec:outside}. In the guiding and central regions, we have
\begin{align}
\psi&\sim H\mathcal{A}(t)+ I\mathcal{B}(t),&t_0<t<t_1,\\
\psi&\sim J\mathcal{A}(t)+ K\mathcal{B}(t),&t_2<t<t_3,\\
\psi&\sim \frac{L\cosh\nu(S(y)-S(x_1))+M\cosh\nu(S(y)-S(x_2))}{q^{1/4}},&x_1<y<x_2,
\label{slot0}
\end{align}
Let us first consider the continuity condition on the normal derivatives at $x=x_1,x_2$:
\begin{align}
\frac{M\sinh\nu S_{21}}{q^{1/4}}&\sim\alpha\sqrt{\eps}\left[ H Ai'\left(-t_1\right)+ I Bi'(-t_1)\right],\\
\frac{L\sinh\nu S_{21}}{q^{1/4}}&\sim-\alpha\sqrt{\eps}\left[ J Ai'\left(-t_2\right)+ K Bi'(-t_2)\right],
\end{align}
with $S_{21}=S(x_2)-S(x_1)$. This directly yields the constants $L$ and $M$ in terms of the other ones. Imposing the continuity of $\psi$ at $y=x_{1,2}$ and treating the remaining interfaces in the same manner as in the previous section, we then obtain the system
\begin{align}
J \mathcal{A}(t_3) + K \mathcal{B}(t_3)&\sim\alpha\sqrt{\eps} \left[ J Ai'\left(-t_3\right)+ K Bi'(-t_3)\right]\left(1+2\ii e^{2\nu S(x_3)}\right), \label{slot1}\\
J \mathcal{A}(t_2) + K \mathcal{B}(t_2)&\sim\frac{\alpha\sqrt{\eps} \left[ H Ai'\left(-t_1\right)+ I Bi'(-t_1)-\cosh(\nu S_{21})\left( J Ai'\left(-t_2\right)+ K Bi'(-t_2)\right)\right]}{\sinh\nu S_{21}},
\label{slot2}\\
H \mathcal{A}(t_1) + I \mathcal{B}(t_1)&\sim\frac{\alpha\sqrt{\eps}\left[ \cosh(\nu S_{21})\left( H Ai'\left(-t_1\right)+ I Bi'(-t_1)\right)-J Ai'\left(-t_2\right)- K Bi'(-t_2)\right]}{\sinh\nu S_{21}} ,
\label{slot3}\\
H \mathcal{A}(t_0) + I \mathcal{B}(t_0)&\sim-\alpha\sqrt{\eps}  \left[H Ai'\left(-t_0\right)+ I Bi'(-t_0) \right],
\label{slot4}
\end{align}
The determinant of this system yields the complex dispersion relation but it is simpler, and more instructive, to solve it by perturbation. Considering the right-hand sides above, two qualitatively distinct situations are directly apparent, depending on the magnitude of $S_{21}$. Indeed, if $S_{21}=O(1)$, then the right-hand sides of Eqs.~(\ref{slot1}) to~(\ref{slot4}) are all small and the two high-index annuli may be regarded as distinct, weakly coupled, waveguides. On the other hand if $\nu S_{21}=O(\sqrt\eps)$, then the four equations are coupled to leading order and we are dealing with a slot waveguide. We postpone the discussion of that situation to Sec.~\ref{sec:slot}. 

In the limit $\sqrt\eps\to0$, Eqs.~(\ref{slot1}) to (\ref{slot4}) read, to leading order, and with the reduced thickness $\hat h_i$ defined above,
\begin{align}
&\left\{\begin{matrix*}[l]
H Ai(-t_1) + I Bi(-t_1)=0\\
H Ai(\hat h_1-t_1) + I Bi(\hat h_1-t_1)=0
\end{matrix*}\right.
&,&
&\left\{\begin{matrix*}[l]
J Ai(-t_3) + K Bi(-t_3)=0\\
J Ai(\hat h_2-t_3) + K Bi(\hat h_2-t_3)=0
\end{matrix*}\right.
\end{align}
so that either
$(H,I,J,K)\sim(1,-Ai(-t_1)/Bi(-t_1),0,0)$ and
\beq
\mathcal{D}(t_1,\hat h_1) =0
\label{t1}
\eeq
or $(H,I,J,K)\sim(0,0,1,-Ai(-t_3)/Bi(-t_3))$ and
\beq
\mathcal{D}(t_3,\hat h_2) =0.
\label{t3}
\eeq

\subsection{Excitation of the exterior waveguide}
Let us assume that Eq.~(\ref{t3}) holds to leading order. Being primarily interested in the shape factor $\sigma$ defined in Eq.~(\ref{shapefactor}), we restrict our attention to the exponentially small terms in Eqs.~(\ref{slot1}-\ref{slot2}) that are associated to radiation losses:
\begin{align}
J Ai(-t_3) + K Bi(-t_3)&\sim2\ii\alpha\sqrt{\eps} \left[ J Ai'\left(-t_3\right)+ K Bi'(-t_3)\right]  e^{2\nu S(x_3)} , \\
J Ai(-t_2) + K Bi(-t_2)&\sim 0.
\end{align}
Comparing with Eqs.~(\ref{shell1}-\ref{shell2}), we notice that the problem is mathematically equivalent at this order to the one-ring problem. Hence, without further calculation, the shape factor $\sigma$ is again given by 
\beq
\sigma=\frac{1}{1-P^2}.
\label{shape:exterior}
\eeq

\subsection{Excitation of the interior waveguide }
Assuming now that Eq.~(\ref{t1}) holds to leading order, we assume the following asymptotic expansion:
\begin{align}
H&=1,\\
I&=-Ai(-\tau)/Bi(-\tau)+\sqrt\eps I_1+\ldots+2\alpha\sqrt{\eps}e^{2\nu S(x_1)}\delta I,\\
t_1&\sim\tau+\sqrt\eps\tau_1+\ldots+2\alpha\sqrt{\eps}e^{2\nu S(x_1)}\delta\tau,\\
J&=\sqrt\eps J_1+\ldots+2\alpha\sqrt{\eps}e^{2\nu S(x_1)}\delta J,\\
K&=\sqrt\eps K_1+\ldots+2\alpha\sqrt{\eps}e^{2\nu S(x_1)}\delta K,
\end{align}
where $\tau$ is now a root of Eq.~(\ref{t1}). Note that the quantities $t_0, t_2$, and $t_3$ are related to $t_1$ and should therefore also be expanded. For instance,
\beq
t_3=\frac{R_3-R_1}{\eps R_1}+\frac{R_3}{R_1}\tau.
\eeq

Also, note that the exponentially small terms above are taken to be proportional to $\exp[2\nu S(x_1)]$, not $\exp[2\nu S(x_3)]$, since light is mostly confined near $r=R_1$, that is, $y=x_1$. Hence, we should evaluate the radiation losses by comparison to a cylinder of radius $R_1$. Considering the $O(\sqrt\eps)$ contributions of Eqs.~(\ref{slot1}) and~(\ref{slot2}), and making use of (\ref{Wronskian}), we find that
\begin{align}
\begin{pmatrix}J_1\\K_1\end{pmatrix}=\frac{\alpha }{\pi\sinh(\nu S_{21})Bi(-\tau_0) \mathcal{D}(t_{3},\hat h_2)}
\begin{pmatrix}Bi(-t_{3})\\-Ai(-t_{3})\end{pmatrix}.
\end{align}
Being primarily interested in the shape factor $\sigma$ appearing in (\ref{shapefactor}), we focus next on the exponentially small terms obtained by substituting the preceding expansions in Eqs.~(\ref{slot1}) to~(\ref{slot4}). In collecting these, one should not omit the contributions that come from expanding the arguments of $\mathcal{A}(t_3)$, $\mathcal{B}(t_3),\ldots$. We obtain
\begin{align}
& Ai(-t_3) \delta J+  Bi(-t_3)\delta K\sim \sqrt{\eps}\left[J_1Ai'(-t_3)+K_1 Bi'(-t_3)\right]\left(\ii e^{2\nu S_{31}}+\frac{R_3}{R_1}\delta\tau\right),\label{tworing:1}\\
& Ai(-t_2) \delta J+Bi(-t_2) \delta K \sim\frac{\alpha\sqrt{\eps}   Bi'(-t_1)  \delta I }{\sinh\nu S_{21}}+ \sqrt{\eps}\left[J_1Ai'(-t_2)+K_1 Bi'(-t_2)\right]\frac{R_2}{R_1}\delta\tau,\\
&Bi(-t_1)\delta I\sim\frac{\alpha\sqrt{\eps}\left[  - Ai'\left(-t_2\right)\delta J- Bi'(-t_2)\delta K\right]}{\sinh\nu S_{21}} +\left[H Ai'(-t_1)+I Bi'(-t_1)\right] \delta\tau,\\
& Bi(-t_0)\delta I \sim \left[H Ai'(-t_0)+I Bi'(-t_0)\right]\frac{R_0}{R_1}\delta\tau, \label{tworing:4}
\end{align}
where $S_{31}=S(x_3)-S(x_1)$. Details of resolution of this last set of equations are  given in Appendix D. Eventually, we obtain
\beq
\delta\tau=-\sigma\times\ii,
\eeq
 with the shape factor
 \begin{align}
 \sigma &= \frac{\left(\eps\alpha^2/\pi^2\right)e^{2\nu S_{31}}}{\left(1-P^2\right)\sinh^2(\nu S_{21}) \mathcal{D}(t_3,\hat h_2)^2}  ,&
 P&=\frac{Bi(-\tau)}{Bi(\hat h_1-\tau)}.
 \label{shape:interior}
 \end{align}
 However, it is more meaningful, in the present case, to compare the above losses to those of a bare shell rather than to those of the full cylinder. Given (\ref{shape:shell}), this yields the new shape factor
 \begin{align}
 \sigma' &= \frac{\left(\eps\alpha^2/\pi^2\right)e^{2\nu S_{31}}}{\sinh^2(\nu S_{21}) \mathcal{D}(t_3,\hat h_2)^2}  .
 \label{shape:interior}
 \end{align}

\begin{figure}
\begin{center}
\includegraphics[width=.45\textwidth]{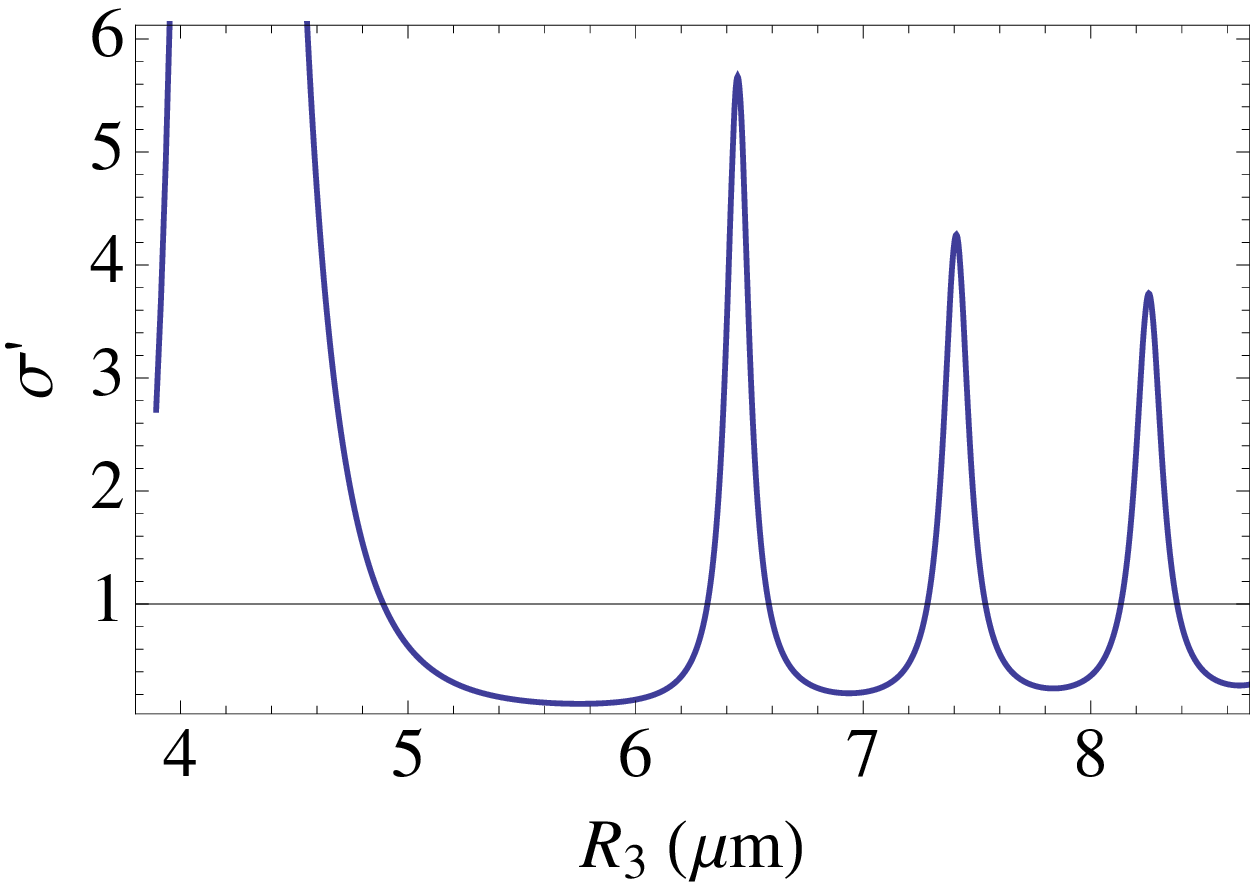}\quad
\includegraphics[width=.45\textwidth]{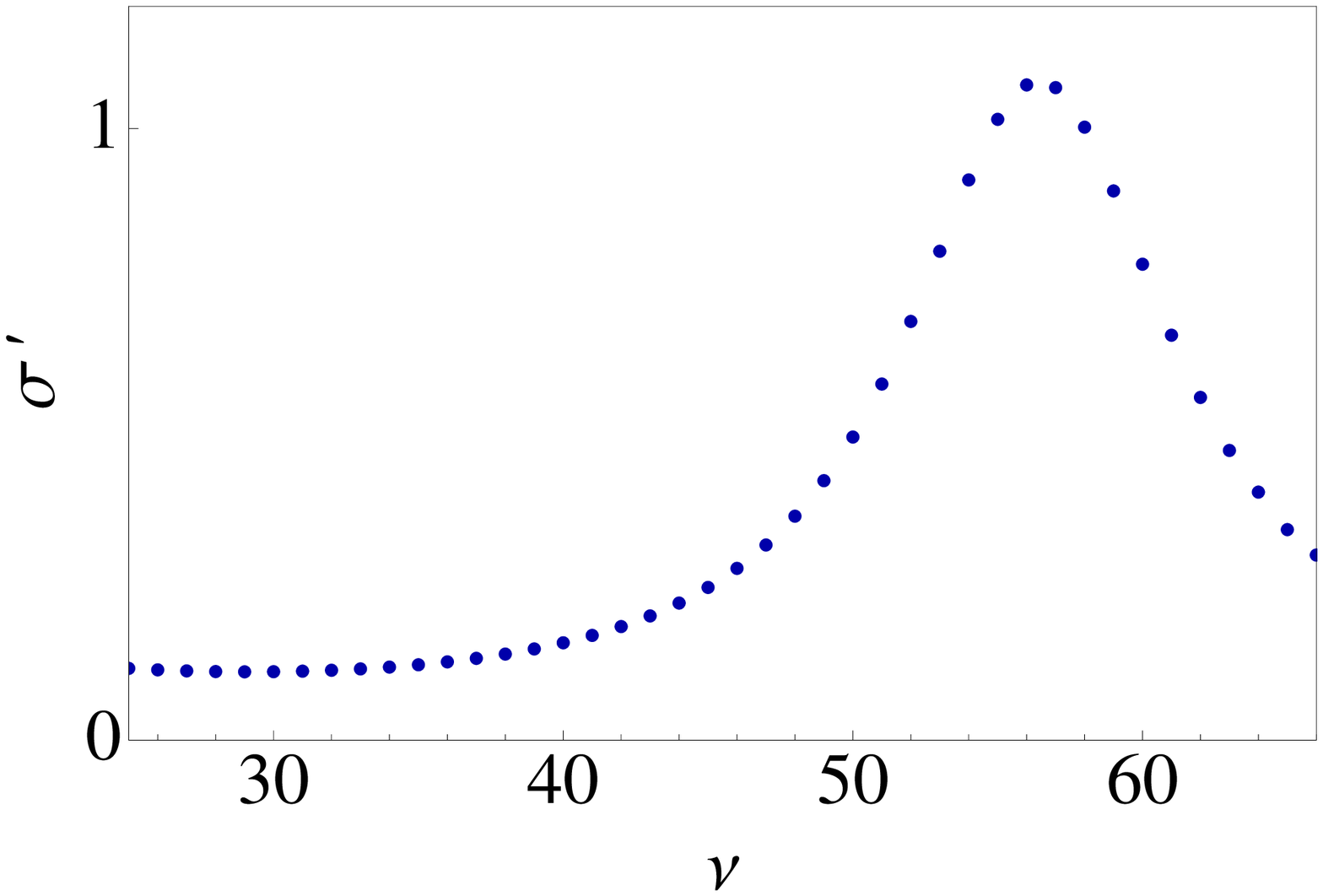}\\
\vspace{.25cm}
\includegraphics[width=.45\textwidth]{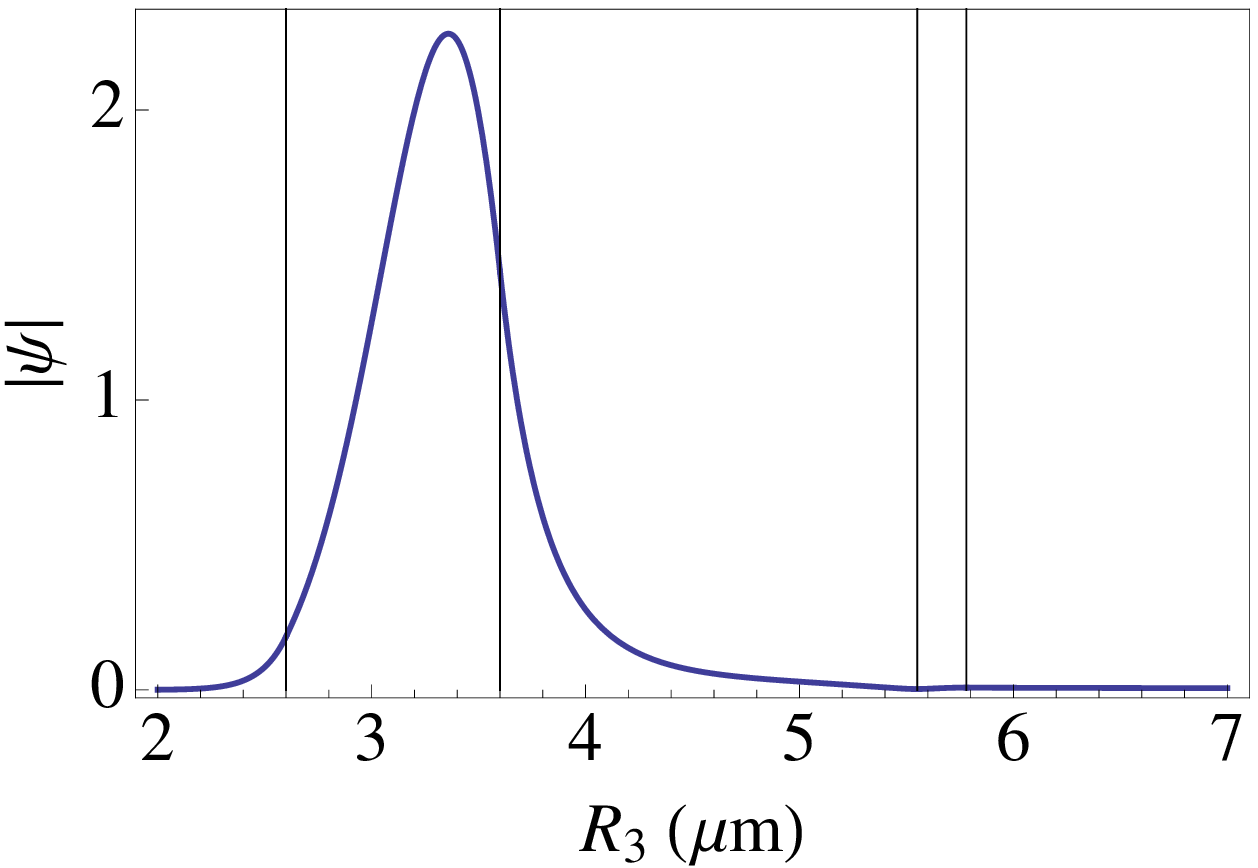}\quad
\includegraphics[width=.45\textwidth]{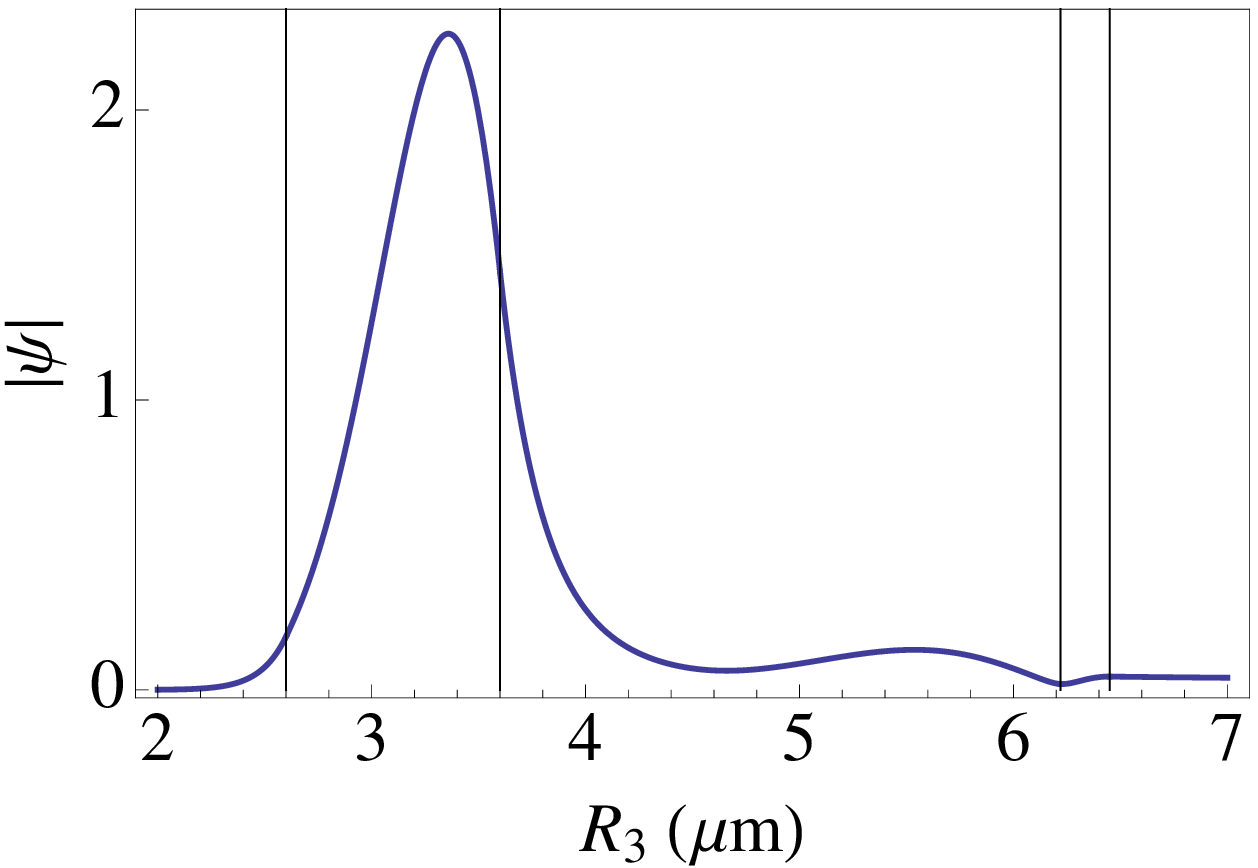}
\caption{Shape factor $\sigma'$ of the two-shell problem, when light is confined in the inner shell. $R_1= 3.6 \mu$m, $h_1= 1 \mu$m, $h_2= 0.23 \mu$m, $n=1.65$ Top left:  $\nu=25$ and variable outer shell position. Top right: $R_3=5.78 \mu$m and variable  $\nu$. Bottom: field distribution corresponding to the minimum  (left) and maximum (right) of $\sigma'$. The vacuum wavelength in both cases is found to be $1.27\mu$m.}
\label{fig:2rings}
\end{center}
\end{figure}

The above expression invites some comments. On the one hand, the $\eps$ factor indicates the possibility of reducing the radiation losses thanks to the presence of the outer shell. On the other hand, $\sigma'$ diverges if  $\mathcal{D}(t_3,\hat h_2)\to0$. Since the vanishing of $\mathcal{D}(t_3,\hat h_2)$ corresponds to sustained oscillation in the exterior shell, this suggests a kind of resonant leakage. The two effects can clearly be seen in Fig.~\ref{fig:2rings}, in which the position of the external shell, and hence the value of $\mathcal{D}(t_3,\hat h_2)$, is varied. In this figure, the roots of the exact two-shell characteristic equation and of the one-shell characteristic equation (see Appendix C) were numerically computed and compared. Both suppression and enhancement of the losses are seen due to the presence of the external shell, as expected from Eq.~(\ref{shape:interior}). The peaks of Fig.~\ref{fig:2rings}  are found to precisely match the resonances of the outer shell alone. The field distribution confirms this: large $\sigma'$ correspond to a significantly larger amplitude in the outer ring compared to when $\sigma'$ is minimum. Further insight is obtained by computing $\sigma'$ as a function of the orbital number $\nu$. Given that $\Delta k/k\approx \Delta \nu/\nu$, where $\Delta k,\Delta\nu$ are variation of $k$ and $\nu$, we see that the resonances are spectrally very broad ($\Delta \nu/\nu\approx0.15$ in Fig.~\ref{fig:2rings}, right). Hence, they do not correspond to resonances between the electromagnetic field and the guiding structure, which are exponentially sharp. Rather, these are structural resonances between the inner and the outer shell.

\section{Slot waveguides}\label{sec:slot}
As is well-known~\cite{Almeida-2004}, slot waveguides are achieved with nanometric low-index gap between high-index layers. This feature arises naturally in our analysis. Indeed,   we found in the preceding section that light is located in  either of the two high-index rings unless $\sinh(\nu S_{21})=O(\sqrt{\eps})$. This requires that the gap $R_2-R_1$ be very small. The main electric field component is the radial one, so the mode is TM and we have $\xi=n^2$ in what follows. We now introduce the normalised gap $\Delta$ as
\begin{multline}
\Delta\equiv\frac{\sinh(\nu S_{21})}{\alpha\sqrt\eps}\sim \frac{\nu S'(x_1)\left(x_2-x_1\right)}{\alpha\sqrt\eps}
\sim 2^{-1/3}\nu^{4/3}\left(n^2-1\right)\frac{R_2-R_1}{R_1}\\
\sim nk(R_2-R_1)\left(nkR_1/2\right)^{1/3} \left(n^2-1\right) 
\end{multline}
and we treat $\Delta$ as an $O(1)$ quantity. Furthermore, we have $t_2-t_1=O(\nu^{-2/3})=O(\eps)$. Hence, we may identify $t_2$ with $t_1$ to leading order and we have, in what follows,
\begin{align}
t_1&\sim t_3-\hat h_2, 
&t_0\sim t_3-\hat h_1-\hat h_2.
\end{align}
The continuity equations are
\begin{align}
J \mathcal{A}(t_3) + K \mathcal{B}(t_3)&\sim\alpha\sqrt{\eps} \left[ J Ai'\left(-t_3\right)+ K Bi'(-t_3)\right]\left(1+2\ii e^{2\nu S(x_3)}\right),\\
J \mathcal{A}(t_1) + K \mathcal{B}(t_1)&\sim\frac{   (H-J) Ai'\left(-t_1\right)+ (I-K) Bi'(-t_1) }{\Delta},
\\
H \mathcal{A}(t_1) + I \mathcal{B}(t_1)&\sim\frac{    (H-J) Ai'\left(-t_1\right)+ (I-K) Bi'(-t_1)}{\Delta} ,
\\
H \mathcal{A}(t_0) + I \mathcal{B}(t_0)&\sim-\alpha\sqrt{\eps}  \left[H Ai'\left(-t_0\right)+ I Bi'(-t_0) \right].
\end{align}
Note that the first equation above is, to leading order, $J Ai(-t_3) + K Bi(-t_3)=0$. Using this piece of information together with Eq.~(\ref{Wronskian}) we may simplify $ J Ai'\left(-t_3\right)+ K Bi'(-t_3)$ by $-J/(\pi Bi(-t_3))$.
In order to reduce the algebra to minimum, we discard all $O(\sqrt\eps)$ terms above, except the one that accounts for radiative losses: 
\begin{align}
J \left[Ai(-t_3)+\frac{2\ii\alpha\sqrt{\eps}  e^{2\nu S} }{\pi Bi(-t_3)}\right] + K Bi(-t_3)&\sim0,\\
J Ai(-t_1) + K Bi(-t_1)&\sim\frac{    (H-J) Ai'\left(-t_1\right)+ (I-K) Bi'(-t_1)}{\Delta} ,
\\
(H-J) Ai(-t_1) + (I-K) Bi(-t_1)&\sim0,
\\
H Ai(-t_0) + I Bi(-t_0)&\sim0.
\end{align}
By successive elimination and repeated use of (\ref{Wronskian}), we find that
\beq
\mathcal{D}(t_3,\hat h_1+\hat h_2)+\pi\Delta \mathcal{D}(t_3, \hat h_2)\mathcal{D}(t_1,\hat h_1 )
+ 
\frac{2\ii\alpha\sqrt{\eps}  e^{2\nu S(x_3)} }{\pi Bi(-t_3)}
\left[Bi(-t_0)+\pi\Delta Bi(-t_1)\mathcal{D}(t_1,\hat h_1 )\right]
\eeq
vanishes. Finally, since the last term above is exponentially small, we have, to a very good approximation, $\pi\Delta \mathcal{D}(t_1,\hat h_1 )\approx -\mathcal{D}(t_3,\hat h_1+\hat h_2)/\mathcal{D}(t_3, \hat h_2)$. As a result, the last term in the characteristic equation may be further manipulated (recalling that $t_2\approx t_1$) to give
\beq
\mathcal{D}(t_3,\hat h_1+\hat h_2)+\pi\Delta \mathcal{D}(t_3, \hat h_2)\mathcal{D}(t_1,\hat h_1 )
- 
\frac{2\ii\alpha\sqrt{\eps}  e^{2\nu S} }{\pi }\frac{\mathcal{D}(t_1,\hat h_1 )}{\mathcal{D}(t_3, \hat h_2)}=0
\eeq
The solution is given by
\beq
t_3\sim\tau -\sigma\times 2\ii\alpha\sqrt{\eps}e^{2\nu S},
\eeq
where $\tau$ satisfies 
\beq
\mathcal{L}(\tau)\equiv \mathcal{D}(\tau,\hat h_1+\hat h_2)+\pi\Delta \mathcal{D}(\tau, \hat h_2)\mathcal{D}(\tau-\hat h_2,\hat h_1 )=0
\eeq
and the shape factor is given by
\beq
\sigma = -\frac1{\mathcal{L}'(\tau)}\frac{\mathcal{D}(\tau-\hat h_2,\hat h_1 )}{\pi\mathcal{D}(\tau, \hat h_2)}.
\label{shape:slot}
\eeq

\section{Discussion}
In this work we have provided an asymptotic framework to study resonances and radiative losses in shells. Although we have limited our attention to one or two concentric shells of the same index of refraction in an homogeneous environment, the approach could be extended to study more general sequences of refractive indexes or to larger sets of shells. Moreover, we expect that the trends that have analytically derived here also hold, broadly, for more intractable geometries, such as toroidal cavities and micro-rings, since their shapes locally conform to portions of spherical or cylindrical shells. Conversely, our analysis does not apply to weakly guiding shell or waveguides, such as optical fibers with a very small contrast between the core and the cladding: $n_2/n_1-1\ll1$. Indeed, in that case the extent of evanescent zone on the outer side of the bend becomes extremely small and the  field almost immediately acquires a radiating  structure, as \emph{e.g.} in Ref.~\cite{Hiremath-2005}. Thus, the WKB expression (\ref{exterior2}), which underlies our analysis, breaks down.

With respect to the study of a full sphere or cylinder, the single-shell problem introduces the thickness $h$ as an additional geometrical parameter. To leading order in the azimuthal number $\nu$, the characteristic equation $\mathcal{D}(\tau,\hat h)=0$ involves a simple combination of Airy functions. It is easy to see that it rapidly becomes equivalent to $Ai(-\tau)=0$, \emph{i.e.} to the characteristic equation for a full sphere as the reduced thickness $\hat h$ increases. In the same way, the coefficient $P$ and $\tilde P$ appearing in expression~(\ref{res_shell}) rapidly tend to zero, eventually leading to the formulas of a full-sphere in Ref.~\cite{Lam-1992}. A simple criterion to treat shells as full-spheres or full-cylindres  is that $\hat h>\alpha^{(j)}$, where $\alpha^{(j)}$ denote the $j^\text{th}$ real root of $Ai(-z)$.

Conversely, let us examine the situation where $\hat h$ is sufficiently small that the shell cannot be considered to be optically equivalent to a full-sphere. Examining the equation $\mathcal{D}(\tau,\hat h)=0$ for small $\hat h$ we find that $\tau$ must be large. Using appropriate asymptotic representation of the Airy function~\cite{Abramowitz}, the $j^\text{th}$ root is found to be
\beq
\tau^{(j)}
\approx \left(j\pi/\hat h\right)^2,
\label{straight}
\eeq
with very good approximation as long as $\hat h \leq 1$ (or, more generally, as long as $\tau^{(j)}$ is much larger than $\hat h$.) In terms of unscaled wavenumber $k$ and thickness $h$, Eq.~(\ref{straight})  is 
\begin{align}
n^2k^2&\approx\beta^2+k_\perp^2,
&\beta&=\nu/R, &k_\perp&=j\pi/h,
\end{align}
which underscores the asymptotic equivalence between a thin shell and a straight waveguide.

Aside from describing the position of the resonances, the motivation of this work was to derive the shape parameter $\sigma$ in the expression of the radiation losses
\begin{align}
k_iR&\sim 
\sigma  \times  \Gamma_0,
&\Gamma_0=\frac{2  e^{2\nu S(x)} }{\xi  \sqrt{n^2-1}} ,
\end{align}
in various configurations. Given the factor $\nu$ in the exponential $\exp2\nu S(x)$ above, it necessary to know  $x$ with at least $O(\nu^{-1})$ accuracy. However, this still leaves an imprecision of order $\nu^{-1/3}$ in the exponential, which decays only slowly with $\nu$. This is what motivated us to derive an expression of $x$ with  $O(\nu^{-4/3})$ precision in Eq.~(\ref{res_shell}). As we have checked, this makes $\Gamma_0$ in numerical agreement in the large-$\nu$ limit with Lam {\it et al.}'s formula~\cite{Lam-1992}: 
\begin{align}
\Gamma_0&\sim\frac{2}{\xi(n^2-1)\left[\nu x\;n_l(\nu x)\right]^2},
&n_l(z)=\sqrt{\frac{\pi}{2z}}Y_{l+1/2}(z).
\end{align}

Regarding the shape factor $\sigma$, we have given an analytical expression (\ref{shape:slot}) for the practically important case of slot waveguides. On the other hand, the case where two concentric shells are too distant to form a slot waveguide is also interesting. Indeed the factor $\eps$ in the shape factor (\ref{shape:interior}) suggests that an external shell can act as a radiation shield for light circulating in the inner shell. It is important to note in this regard that the outer shell can be located in the evanescent zone. Hence, this radiation shielding can not be interpreted as resulting from some sort of reflection by the outer shell of light radially emitted by the inner shell.  In addition, the expression for $\sigma$ in (\ref{shape:interior})  also reveals that radiation losses can be enhanced if $\mathcal{D}(t_3,\hat h_2)$ is small, \emph{i.e.} if the inner and outer shells are in resonance. In that case, instead of shielding radiation, the outer shell acts as a resonant escape channel. Our analysis, carried out in the large-$\nu$ limit, is well confirmed even for the moderately large values of $\nu=25$ in Fig.~\ref{fig:2rings}. It should be noted that the radiation losses can be modulated by a factor of 50 in that figure through the location of the outer ring. Even larger ratios between the largest and lowest $\sigma$ can be achieved for larger $\nu$, \textit{i.e.} larger rings or shorter wavelengths. An important conclusion is that the concentric circular structure offers a simple and powerful way to control the $Q$-factor of WGM resonators, with interesting perspectives of applications for both active and passive devices.

\section*{Funding}
GK is a research associate with the Fund for Scientific Research - FNRS. This project has received funding from the European Union's Horizon 2020 research and innovation programme under grant agreement No 634928.

\appendix

\section*{Appendix A: WKB analysis outside the shell}
\setcounter{equation}{0}
\renewcommand{\theequation}{A{\arabic{equation}}}
In equation (\ref{Helm:1}), let us substitute
\begin{align}
\psi &=  \phi(y)e^{\nu S(y)},
& S'(y)&=\pm\sqrt{q(y)}.
\label{WKBansatz}
\end{align}
The equation for $\phi$ becomes
\beq
2S'\phi'+\left(S''+\frac{S'}y\right)\phi=-\frac1{\nu y}\left(y \phi'\right)'
\eeq
and can be solved by successive approximations in powers of $\nu^{-1}$. With a bit of hindsight, we may simplify it by writing
\begin{align}
\phi(y)&=\frac{f(y)}{\sqrt{yS'(y)}}, &f'(y)&=-\frac{\left(y \phi'(y)\right)'}{2\nu \sqrt{yS'(y)}}.
\end{align}
Then, expanding $f$ as $f\sim f_0+\nu^{-1}f_1+\nu^{-2}f_2+\ldots,$ we obtain the recursion
\begin{align}
f_{j+1}'(y)&=-\frac{\left(y \phi_j'(y)\right)'}{2\sqrt{yS'(y)}}, &\phi_{j}&=\frac{f_{j}}{\sqrt{yS'(y)}}.
\end{align}
The solution is thus asymptotically given by
\beq
\psi_\pm\sim\text{const.}\times\frac{e^{\pm\nu S(y) } }{\left|1-y^2\right|^{1/4}}\left(1+\sum_{j\ge1} \nu^{-j}f_j\right).
\label{WKB}
\eeq

\section*{Appendix B: asymptotic resolution of the one-shell characteristic equation}
\setcounter{equation}{0}
\renewcommand{\theequation}{B{\arabic{equation}}}

Let us solve Eqs.~(\ref{shell1}) and (\ref{shell2}) with the aid of the following expansions
\begin{align}
H&=1,\\
I&\sim i_0+  \eps^{1/2} i_1+ \eps i_2+\ldots+2\alpha\sqrt{\eps}e^{2\nu S(x)}\delta i,\\
t_1&\sim \tau +  \eps^{1/2} \tau_1+\eps \tau_2+\ldots+2\alpha\sqrt{\eps}e^{2\nu S(x)}\delta \tau,\\
t_0&=\left(1-\eps\hat h\right)t_1-\hat h.
\end{align}
In these expansions for $I$ and $t_1$, the last terms are exponentially small and smaller than any finite power of $\eps$ as $\nu\to\infty$.  Substituting them in  Eqs.~(\ref{shell1}) and (\ref{shell2}),  we find  at leading order  that $i_0=-Ai(-\tau)/Bi(-\tau)$, and that $\tau$ must satisfy $\mathcal{D}(\tau,\hat h)=0$. Equations~(\ref{shell1}) and (\ref{shell2}) are sufficiently precise that we may pursue the analysis for two more orders in the $\sqrt\eps$-expansion. As this is a standard regular perturbation problem, we omit the details for brevity. We find
\begin{align}
\tau_1&=-\frac{1+P^2}{1-P^2}\alpha,& 
\tau_2&=\frac{3\tau^2}{10}-\frac{P^2}{1-P^2}\left(\frac{3\hat h^2+4\hat h\tau}{10} 
-4\alpha^2\frac{\left(1-P\tilde P\right)Bi'(\hat h-\tau)}{\left(1-P^2\right)^2 Bi(\hat h-\tau)}
\right)
\end{align}
where
\begin{align}
P&=\frac{Bi(-\tau)}{Bi(\hat h -\tau)} , &\tilde P&=\frac{Bi'(-\tau)}{Bi'(\hat h -\tau)}.
\end{align}
Finally, it remains to compute $\delta i$ and $\delta \tau$. Eqs.~(\ref{shell1}) and (\ref{shell2}) yield
\begin{align}
\delta i  Bi(-\tau)& =\left[H Ai'(-\tau)+IBi'(-\tau)\right]\left( \ii+\delta\tau\right),\\
\delta i  Bi(\hat h-\tau)& =\left[H Ai'(\hat h-\tau)+IBi'(\hat h-\tau)\right] \delta\tau ,
\label{shell4}
\end{align}
Using the fact that $I\sim-H Ai(-\tau)/Bi(-\tau)=-H Ai(\hat h-\tau)/Bi(\hat h-\tau)$ and the identity
\beq
Ai(z)Bi'(z)-Ai'(z)Bi(z)=1/\pi,
\label{Wronskian}
\eeq
one easily obtains, with $P$ defined above,
\beq
\delta\tau=-\frac1{1-P^2}\times\ii.
\eeq
Combining the definitions (\ref{def:y}), (\ref{def:t}), (\ref{epsilon:1}), and (\ref{def:alpha}) of $y$, $t$, $\eps$ and $\alpha$, respectively,
\beq
k_i R  \sim 
\frac{1}{1-P^2}\times  \frac{2  e^{2\nu S} }{\xi  \sqrt{n^2-1}}.
\eeq
Hence, the imaginary part of $\delta\tau$ directly yield the shape factor, $\sigma$.

\section*{Appendix C: Exact characteristic equations}
\setcounter{equation}{0}
\renewcommand{\theequation}{C{\arabic{equation}}}
The characteristic equation for the cylinder is
\beq
n J'_\nu(nkR_1)H^{(1)}_\nu(kR_1)-J_\nu(nkR_1)H'^{(1)}_\nu(kR_1)=0,
\eeq
where $H^{(1)}_\nu$ is the Hankel function of the first kind. For the one-ring problem, the characteristic equation is
\beq
\left|\begin{matrix}
J_\nu(kR_0)&-J_\nu(nkR_0)&-Y_\nu(nkR_0)&0\\
J'_\nu(kR_0)&-nJ'_\nu(nkR_0)&-nY'_\nu(nkR_0)&0\\
0& J_\nu(nkR_1)&Y_\nu(nkR_1)&-H^{(1)}_\nu(kR_1)\\
0& nJ'_\nu(nkR_1)&nY'_\nu(nkR_1)&-H'^{(1)}_\nu(kR_1)
\end{matrix}
\right|=0.
\eeq
Finally, for the two-ring problem, the determinant is

\beq
{\footnotesize 
\left|\begin{matrix}
J_\nu(kR_0)&-J_\nu(nkR_0)&-Y_\nu(nkR_0)&0&0&0&0&0\\
J'_\nu(kR_0)&-nJ'_\nu(nkR_0)&-nY'_\nu(nkR_0)&0&0&0&0&0\\
0& J_\nu(nkR_1)&Y_\nu(nkR_1)&-J_\nu(kR_1)&-Y_\nu(kR_1)&0&0&0\\
0& nJ'_\nu(nkR_1)&nY'_\nu(nkR_1)&-J'_\nu(kR_1)&-Y'_\nu(kR_1)&0&0&0\\
0&0&0&J_\nu(kR_2)&Y_\nu(nkR_2)&-J_\nu(nkR_2)&-Y_\nu(nkR_2)&0\\
0&0&0&J'\nu(kR_2)&Y'_\nu(nkR_2)&-nJ'_\nu(nkR_2)&-nY'_\nu(nkR_2)&0\\
0&0&0&0&0&J_\nu(kR_3)&Y_\nu(nkR_3)&-H^{(1)}_\nu(kR_3)\\
0&0&0&0&0&nJ'_\nu(kR_3)&nY'_\nu(nkR_3)&-H'^{(1)}_\nu(kR_3)\\
\end{matrix}
\right|.
}
\eeq

\section*{Appendix D: asymptotic resolution of the two-shell characteristic equations}
\setcounter{equation}{0}
\renewcommand{\theequation}{D{\arabic{equation}}}
Equations~(\ref{tworing:1}) to (\ref{tworing:4}) can be simplified by noting that, to leading order, we have simultaneously, $H=1$,  $I\sim -Ai(-t_0)/Bi(-t_0)\sim -Ai(-t_1)/Bi(-t1)$, $I\sim -Ai(-t_0)/Bi(-t_0)\sim -Ai(-t_1)/Bi(-t1)$, $K_1=-Ai(t_3)J_1/Bi(-t_3)$. Then, using the identity~(\ref{Wronskian}), we obtain
\begin{align}
 Ai(-t_3) \delta J+  Bi(-t_3)\delta K &\sim - \frac{\sqrt{\eps}J_1}{\pi Bi(-t_3)}\left(\ii e^{2\nu S_{31}}+\frac{R_3}{R_1}\delta\tau\right),\\
 Ai(-t_2) \delta J+Bi(-t_2) \delta K &\sim\frac{\alpha\sqrt{\eps}   Bi'(-t_1)  \delta I }{\sinh\nu S_{21}}+ \sqrt{\eps}\left[J_1Ai'(-t_2)+K_1 Bi'(-t_2)\right]\frac{R_2}{R_1}\delta\tau,\\
Bi(-t_1)\delta I &\sim\frac{\alpha\sqrt{\eps}\left[  - Ai'\left(-t_2\right)\delta J- Bi'(-t_2)\delta K\right]}{\sinh\nu S_{21}} -\frac{\delta\tau}{\pi Bi(-t_1)},\label{deltaI}\\
 Bi(-t_0)\delta I &\sim -\frac{(R_0/R_1)\delta\tau}{\pi Bi(-t_0)},
\end{align}
From the assumption that $\hat h_1=O(1)$, we have $R_0/R_1\sim1+O(\eps)$. Furthermore, on closer inspection, we notice that $\delta J,\delta K=O(\sqrt\eps)$ while $\delta I,\delta\tau=O(\eps)$. Hence
\begin{align}
 Ai(-t_3) \delta J+  Bi(-t_3)\delta K&\sim \frac{-\ii \sqrt{\eps}J_1}{\pi Bi(-t_3)}  e^{2\nu S_{31}} ,\\
 Ai(-t_2) \delta J+Bi(-t_2) \delta K &\sim0, \label{deltaK}\\
Bi(-t_1)\delta I &\sim\frac{\alpha\sqrt{\eps}\delta J}{\pi \sinh\nu S_{21}} -\frac{\delta\tau}{\pi Bi(-t_1)},\label{deltaI2}\\
 Bi(-t_0)\delta I &\sim -\frac{\delta\tau}{\pi Bi(-t_0)},
\end{align}
where (\ref{deltaK}) was used to simplify (\ref{deltaI}). Successive elimination eventually yields Eq.~(\ref{shape:interior}).

\end{document}